\documentclass[usenatbib]{mn2e}
\usepackage{graphicx}
\usepackage[usenames]{color}
\usepackage{times}
\usepackage{rotating}
\def\gtsim{\lower.5ex\hbox{$\; \buildrel > \over \sim \;$}}
\def\ltsim{\lower.5ex\hbox{$\; \buildrel < \over \sim \;$}}

\title[Galaxy Surface Brightness]{Galaxy And Mass Assembly (GAMA): The Dependence of Star Formation on Surface Brightness in Low Redshift Galaxies}
\author[S. Phillipps et al.]{S. Phillipps$^1$, S. Bellstedt$^{2}$, M.N. Bremer$^1$, 
R. De Propris$^{3,4}$, P.A. James$^5$,
\and S. Casura$^6$, J. Liske$^6$
and B.W.Holwerda$^7$\\
$^1$Astrophysics Group, School of Physics, University of Bristol, Tyndall Avenue, Bristol, BS8 1TL, UK\\
$^{2}$ICRAR, University of Western Australia, 35 Stirling Highway, Crawley, WA6009, Australia\\
$^3$Finnish Centre for Astronomy with ESO, University of Turku, Finland, Vesilinnantie 5, FI-21400, Turku, Finland\\
$^4$Department of Physics and Astronomy, Botswana International University of Science and Technology, Private Bag 16, Palapye, Botswana\\
$^5$Astrophysics Research Institute, Liverpool John Moores University, IC2, Liverpool Science Park, 146 Brownlow Hill, Liverpool L3 5RF, UK\\
$^{6}$Hamburger Sternwarte, Universit\"{a}t Hamburg, Gojenbergsweg 112, 21029 Hamburg, Germany\\
$^7$University of Louisville, Department of Physics and Astronomy, 102 Natural Science Building, Louisville KY 40292, USA
}
\begin{document}

\date{Accepted . Received ; in original form }

\pagerange{\pageref{firstpage}--\pageref{lastpage}} \pubyear{}

\maketitle

\label{firstpage}

\begin{abstract}
The star formation rate in galaxies is well known to correlate with stellar mass (the `star-forming main sequence'). Here we extend this further to explore any additional dependence on galaxy surface brightness, a proxy for stellar mass surface density. We use a large sample of low redshift ($z \leq 0.08$) galaxies from the GAMA survey which have both SED derived star formation rates and photometric bulge-disc decompositions, the latter providing measures of disc surface brightness and disc masses. Using two samples, one of galaxies fitted by a single component with S\'{e}rsic index below 2 and one of the discs from two-component fits, we find that once the overall mass dependence of star formation rate is accounted for, there is no evidence in either sample for a further dependence on stellar surface density. 

\end{abstract}


\section{Introduction}
Star formation and its history are key elements in the evolution of any galaxy. Elucidating the factors which determine the star formation rate (SFR) in a given galaxy at any particular time is therefore key to our understanding of global galaxy evolution. 

It is well established that the SFR depends on the stellar mass ($M_*$) of a galaxy; galaxies which have significant ongoing star formation follow a `star forming main sequence' \citep[][and many others]{Noeske2007,Speagle2014} 
in a plot of SFR against $M_*$ . As the relationship is not linear, there is a corresponding correlation between specific star formation rate (the SFR per unit mass; sSFR) and mass \citep[e.g.][and references therein]{Davies2016,Davies2019}. 

A number of authors \citep[e.g.][]{Ellison2008, Mannucci2010, Lopez2013, Salim2014, Telford2016} have extended this to explore a possible three-way correlation between galaxy (gas phase) metallicity, mass and SFR (at both low and high redshift), with some finding that at given mass the metallicity is higher in lower SFR galaxies, while others argue that the variation in metallicity is entirely a function of mass \citep[e.g.][]{Sanchez2013,Sanders2015}. Most recently, \cite{Thorne2022} do not see a strong dependence of the mass-metallicity relation on SFR. On the other hand, \cite{Peng2015} showed a dependence of stellar metallicity on SFR, which they took as evidence for the strangulation model of star formation reduction. Recent works by \cite{Curti2020} and \cite{Bellstedt2021} summarise the status of fits to a mass-metallicity-SFR plane, the latter finding that the residual effect of SFR on metallicity (or vice versa) is lessened at small look-back-times (low $z$) in galaxies such as those studied here.

Given that SFR depends on the mass of stars already formed, another factor which has been suggested to play a role in the star formation history (SFH) is the surface brightness (SB), or more physically the stellar surface density, of the star forming disc \citep[e.g.][]{BelldJ2000}. Generically, this might be expected if the gravitational instability which leads to the star formation depends on the disc dynamical time or angular velocity \citep[e.g., as summarised in][]{Li2006}, which for a given mass are obviously related to the radius and hence surface density. Similarly, self-regulating star formation theories suggest that at fixed gas density, sSFR should depend on stellar or total mass surface density \citep{Dopita1985, Matteucci1989, Kennicutt2012}, while stability of gas discs in general depends on both the gaseous and stellar components \citep[e.g.][]{JogSolomon,Elmegreen1995,Dalcanton2004,Martig2009}. 

However, much of the work and interest in this possible dependence has centred specifically on the class of low SB galaxies \citep[LSBGs; e.g.][]{Bell2000, Boissier2008}, often with rather small samples of objects. Less attention has been paid to the effects of variations in SB amongst the full population of spiral discs (i.e. not just LSBGs), and the use of the large galaxy samples which have become available in recent years.

In the current paper we use data from the GAMA (Galaxy And Mass Assembly) survey \citep{Driver2011, Driver2022} to explore the connections between SFR and disc SB in spiral galaxies. Since we wish to concentrate on discs (assumed to provide the dominant contribution to the present day star formation) we utilise the most recent photometric bulge-disc decompositions of GAMA galaxies presented by \cite{Casura2022}. Our goal is to examine the correlations between the SFR and stellar surface density, if any exist, in the large statistically complete GAMA sample which enjoys both multi-wavelength SFR estimates and uniformly determined profile fits.

Section 2 describes the GAMA spectroscopic and associated photometric data and the analysis of the radial SB profiles. Section 3 presents our main results on the variation of SFR with SB (and by proxy stellar surface density) and Section 4 discusses the results. 

Where required we use a {\it Planck} 2015 cosmology (Planck Collaboration XIII 2016) with $H_0 = 67.8$km~s$^{-1}$Mpc$^{-1}$, $\Omega_m = 0.308$, $\Omega_{\Lambda} = 0.692$, as in \cite{Bellstedt2020b} from where we obtain our stellar masses and SFRs.

\section{Data and Sample}

Our data arise from the spectroscopic GAMA survey and its allied multi-wavelength photometric data sets. GAMA was a large, highly complete spectroscopic survey carried out with the AAOmega spectrograph on the Anglo-Australian Telescope which obtained redshifts for around 300,000 galaxies in three equatorial fields (G09, G12 and G15) and two southern fields (G02 and G23). Data releases are described in \cite{Driver2011}, \cite{Liske2015}, \cite{Baldry2018} and \cite{Driver2022}. 

The original optical photometry (and therefore selection) for GAMA was obtained from SDSS \citep[specifically their DR8;][]{Aihara2011} but in the latest releases \citep{Driver2022} this has been replaced by data from KiDS \citep[the Kilo-Degree Survey;][]{deJong2013}. KiDS is a
wide-field imaging survey of the Southern sky in the optical broad-band filters $u, g, r, i$ carried out using the VLT Survey Telescope (VST) at the ESO Paranal Observatory. The VISTA Kilo-degree Infrared Galaxy (VIKING) survey provides the corresponding near-infrared data in the $Z, Y, J, H, Ks$ bands \citep{Edge2013, Wright2019}. The GAMA II equatorial survey regions have been covered as of KiDS DR3.0 \citep{deJong2017}. Using this photometry \citep[see][]{Bellstedt2020a}, \cite{Driver2022} demonstrate that the GAMA spectroscopic sample in the equatorial regions is 95\% complete to a KiDS $r-$band magnitude limit of $r=19.65$.

In order to specifically concentrate on star forming galaxy discs we require bulge/disc decomposition of two-component systems where appropriate, i.e. for the earlier type spirals. This limits us to relatively nearby (i.e. sufficiently well resolved) galaxies and we take the overall sample from \cite{Casura2022} which contains 13,096 galaxies out to $z=0.08$.

\cite{Casura2022} fit both single S\'{e}rsic and two-component (point source or bulge plus exponential disc) SB profiles to the KiDS images by using the Bayesian two-dimensional fitting programme ProFit \citep{Robotham2017}. In what follows we utilise their choice of `recommended' profile type, viz. their parameter JOINT-NCOMP; 1 (single component, with fitted S\'{e}rsic index $n$), 1.5 (unresolved point source + exponential) or 2 (bulge with fitted $n$ + exponential). See \cite{Casura2022}, section 3.3.2, for the discussion of this choice. A total of 8725 galaxies are well fit by one of these three options. 

\begin{figure}
\includegraphics[width=\linewidth]{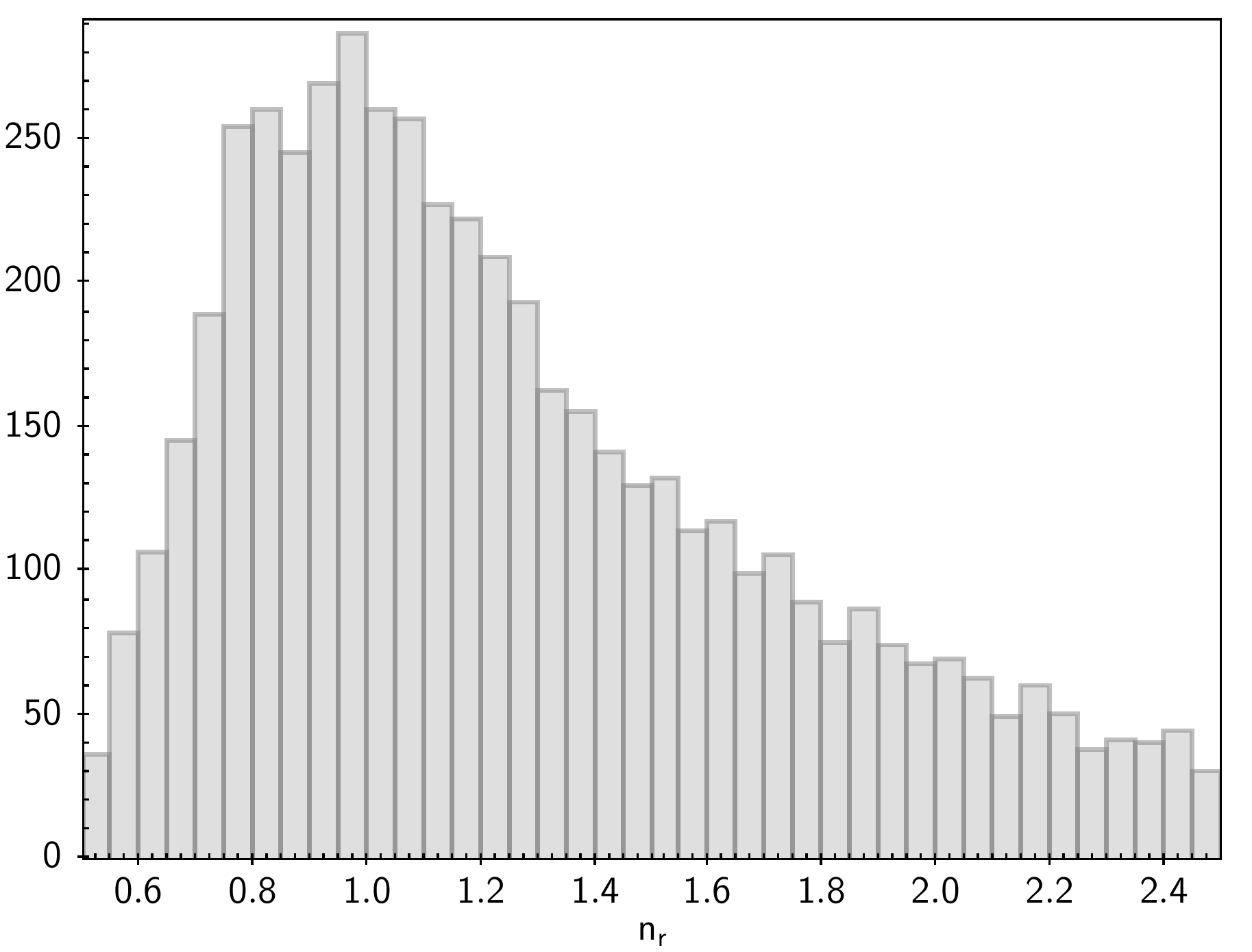}
\caption{Distribution of $r-$band S\'{e}rsic indices $n_r$ (excluding those above $n_r=2.5$) for the single component fits.
}
\label{onesersic}
\end{figure}

Of the galaxies best fit by a single S\'{e}rsic component, we choose those with $0.5 \leq n \leq 2.0$ as our `pure disc' galaxy subsample. We will refer to these 4572 objects as Sample 1. The distribution of their $r-$band S\'{e}rsic indices is shown in Fig. \ref{onesersic}. Changing the limit for discs to $n=2.5$ has no significant effect on our subsequent results. Where morphological classifications exist for these galaxies \citep[at $z < 0.06$;][]{Moffett2016}, as expected, these objects are nearly all classed as late type spirals, with a subset of `little blue spheroids' \citep[i.e., star forming dwarfs; see][]{Moffett2019} at the low mass end. Errors given in the Casura et al. catalogue are typically less than 2\% in the half-light radius $R_e$ and 0.02 magnitudes in $r$. 

\begin{figure}
\includegraphics[width=\linewidth]{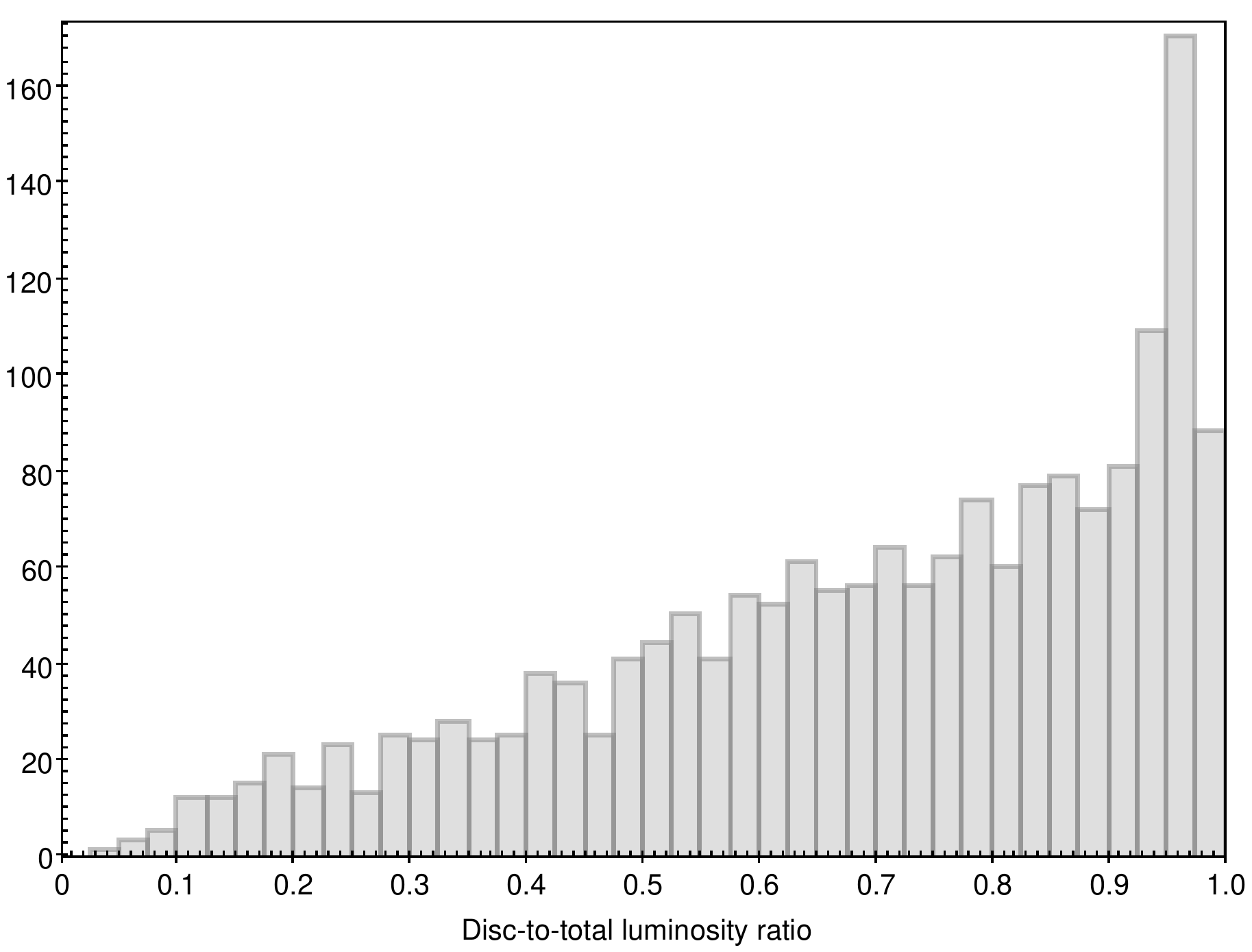}
\caption{Distribution of $r-$band disc-to-total luminosity fractions for the galaxies fitted by two components.
}
\label{twosersic}
\end{figure}

For the two-component fits (types 1.5 and 2), we simply use the disc component parameters with no further restrictions (these necessarily have $n=1$). Fig \ref{twosersic} shows the disc-to-total luminosity ratios in the $r-$band for these 1790 galaxies, henceforth Sample 2. The remainder of the light is ascribed to a generic central `bulge', whether resolved or not; 1405 are disc dominated with disc fraction above 0.5. Again quoted errors are typically less than 0.02 magnitudes in $r$ and 2\% in $R_e$.

\section{Correlations Between Surface Brightness and Star Formation}

\subsection{Surface Brightness}

For simplicity (and generality), we define our characteristic surface brightness parameter $\mu$ to be the mean SB (in magnitudes per square arc second) inside the effective (half-light) radius, i.e.

$\mu \equiv \bar{\mu_e} = (m + 0.75) + 2.5~$log$(\pi R_e^2) = m + 5~$log$R_e$ + 2.0\\
where $m$ is the total apparent magnitude in a given band and $R_e$ is the corresponding effective radius, as given by the fitting procedure. Note that for Sample 1, these refer to the whole galaxy while for Sample 2 they are for the disc component.

\begin{figure}
\includegraphics[width=\linewidth]{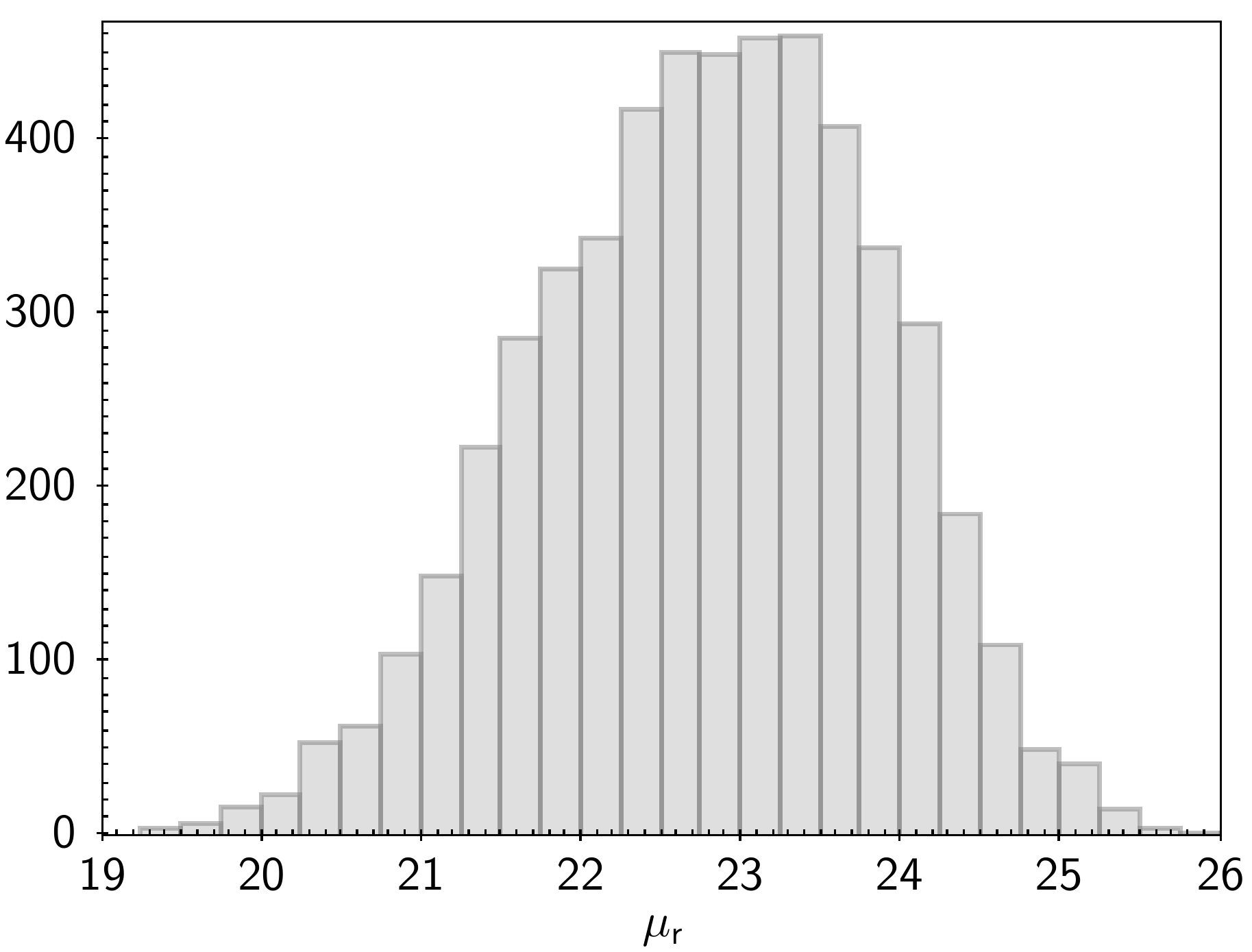}
\caption{Distribution of $r-$band characteristic SB (in magnitudes per square arc second) for the galaxies fitted by a single component and selected as discs.
}
\label{onesersicsb}
\end{figure}

\begin{figure}
\includegraphics[width=\linewidth]{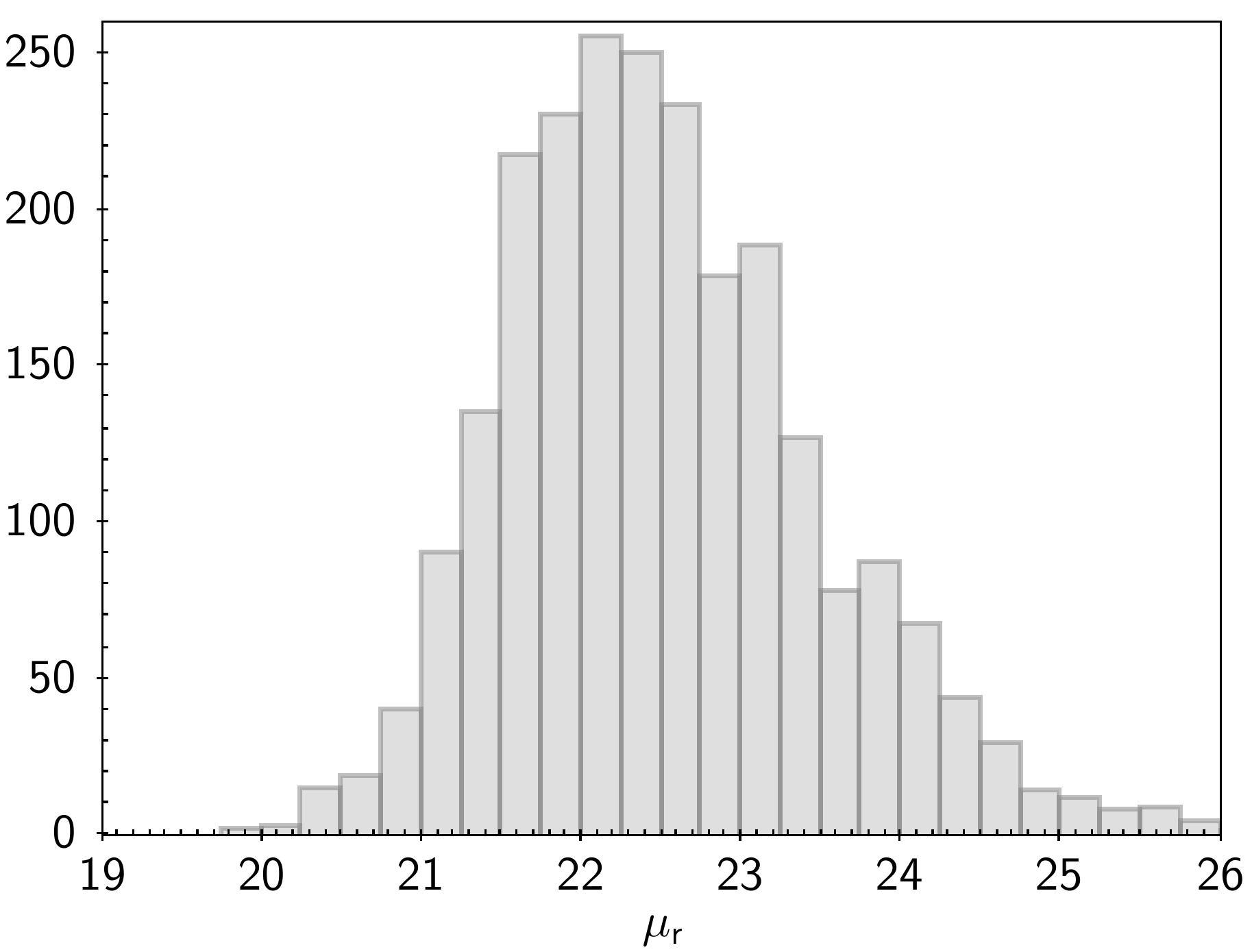}
\caption{Distribution of $r-$band characteristic SB (in magnitudes per square arc second) of the disc component for the galaxies fitted by two components. 
}
\label{twosersicsb}
\end{figure}

Fig. \ref{onesersicsb} and Fig. \ref{twosersicsb} show the overall distributions of $\mu$ (in the $r-$band in this case) for the two samples. Given the quoted errors on the magnitude and effective radius, errors in $\mu_r$ are typically less than 0.05 magnitudes per square arc second.

The simple prescription avoids the details of the galaxy and allows a comparison across galaxies with different values of the S\'{e}rsic index. Nevertheless, we can see that for a perfect $n=1$ face-on exponential disc, for instance, $\mu$ is trivially related to the SB at $R_e$, $\mu_e$, and to the central SB, $\mu_0$, through 

$\mu = \bar{\mu_e} = \mu_e - 0.70 = \mu_0 + 1.12$\\
\citep{Graham2005}.

Since we have used $ \pi R_e^2$ in our definition, rather than the area of an elliptical image, $\pi ab$ (= $\pi R_e^2 (b/a)$, where $b/a$ is the axis ratio), we also automatically correct for the increase in apparent surface density with inclination of the disc (barring any small change in apparent $r_e$ with inclination). We have not here made any correction for dust extinction in the galaxy or its variation with inclination; we consider these later. We do, though, make the small $(1+z)^4$ correction for cosmological dimming.

While statistics of SB measurements are typically reported at blue wavelengths \citep{Freeman1970, Phillipps1987, Oneil2000}, here we concentrate on the $r-$band SB. This more accurately portrays surface mass density and avoids, or at least reduces, correlations forced purely by recently formed bright blue stars contributing a significant amount of the light. We also have $i-$band profiles, where the mass-to-light ratio is even less dependent on the SFR, and we use these as a consistency check, although the errors on the profile fit parameters are slightly larger than for the $r-$band. We initially use our single S\'{e}rsic sample (with $n \leq 2$) and then repeat the analysis with the discs from the two-component fits.

\subsection{Star Formation Rates}

There are numerous options for determining the SFR of the GAMA galaxies \citep{Davies2016}. Here we make use of the SFRs derived by \cite{Bellstedt2020b} using the ProSpect software \citep{Robotham2020} to fit a star formation history model to the multi-wavelength broad-band photometry. The data used to form the SED span the FUV and NUV from GALEX, $ugri$ from the VST, $ZYJHK_s$ from VISTA, $W1$ to $W4$ from WISE and P100 to S500 from Herschel.\footnote{GAMA Data Management Unit GAMAKidsVikingFIRv01; Driver et al. (2016). Note that not all galaxies have data beyond $W2$.} The code then obtains a fit to the SED using a parameterised SFH and a physically motivated metallicity evolution.  ProSpect also provides the present day stellar mass of the galaxy. (Using the MAGPHYs \citep{daCunha2008} derived SFRs from \cite{Driver2016} and \cite{Wright2016} does not materially affect our results). From the ProSpect fits, errors are typically 15\%, i.e. 0.06 dex, in SFR (except at very low values) and 12\% (0.05 dex) in stellar mass.

\begin{figure}
\includegraphics[width=\linewidth]{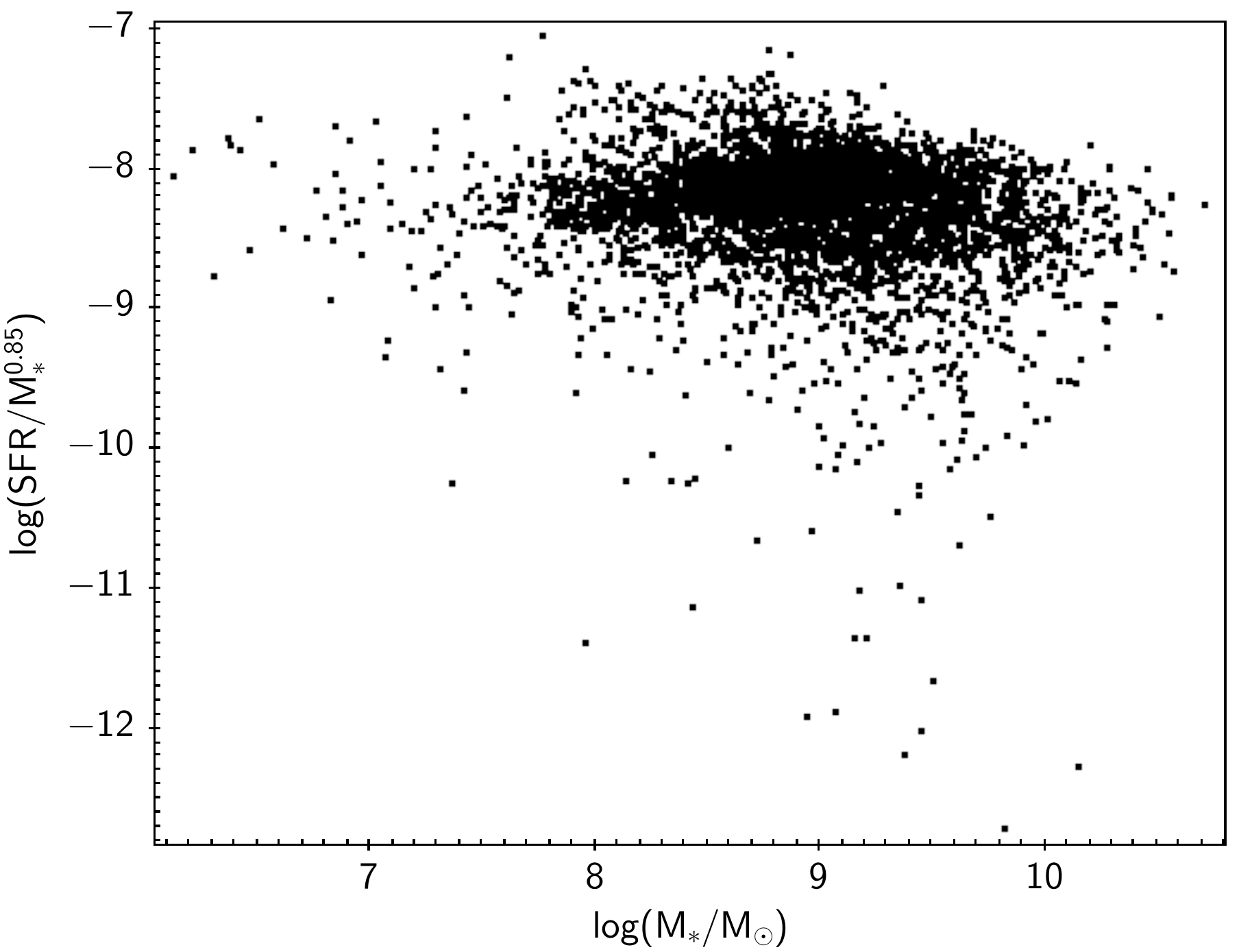}
\caption{Variation of the mass-corrected star formation rate parameter, m$_1$SFR (in units of $M_{\odot}^{0.15}$/yr), with stellar mass (in solar masses) for Sample 1.
}
\label{onesersicmsfrm}
\end{figure}

We first consider the total SFR. Since it is well known that SFR depends strongly on stellar mass, we account for this before looking at the SB. Specifically we determine the slope $m$ of a linear fit to log(SFR) vs log($M_*$), via an unweighted least squares fit and use this to obtain a new parameter 

log(mSFR) = log(SFR) - $m$ log($M_*$)\\ which accounts for the overall mass dependence. For our single S\'{e}rsic sample, Sample 1, $m_1=0.85$, i.e. m$_1$SFR = SFR/$M_*^{0.85}$. For reference, Fig. \ref{onesersicmsfrm} shows the `main sequence of star formation' for Sample 1 in these terms; by design it shows no mass dependence. Note that the parameter m$_1$SFR has the units of $M_{\odot}^{0.15}$/yr. It is not the same as using the specific star formation rate (sSFR), since the slope is slightly sub-linear, i.e. sSFR decreases with mass.  This result - for single-component disc-like galaxies only - is counter to the assertion of \cite{Abramson2014} that the slope of the usual main sequence of star formation deviates from unity only because of lower disc fractions at higher overall mass.

Notice that there are some quiescent galaxies, well below the main sequence, at all masses (though more so at high mass). These quiescent galaxies often appear as outliers in subsequent plots, but removing them does not alter any of our conclusions.  (Physically, the outliers tend to be preferentially towards higher masses, higher surface brightnesses and larger bulge-to-disc ratios compared to the overall sample). As noted by \cite{Eales2017} and \cite{Oemler2017} \citep[though see][]{Salim2014,Holwerda2022}, there is no obvious `valley' between main sequence and quiescent galaxies when plotting SFR, as opposed to the case when using optical or UV-optical colour \citep[e.g.,][for GAMA]{Bremer2018}. Errors in m$_1$SFR are typically less than 0.1 dex except at the very low levels.

\begin{figure}
\includegraphics[width=\linewidth]{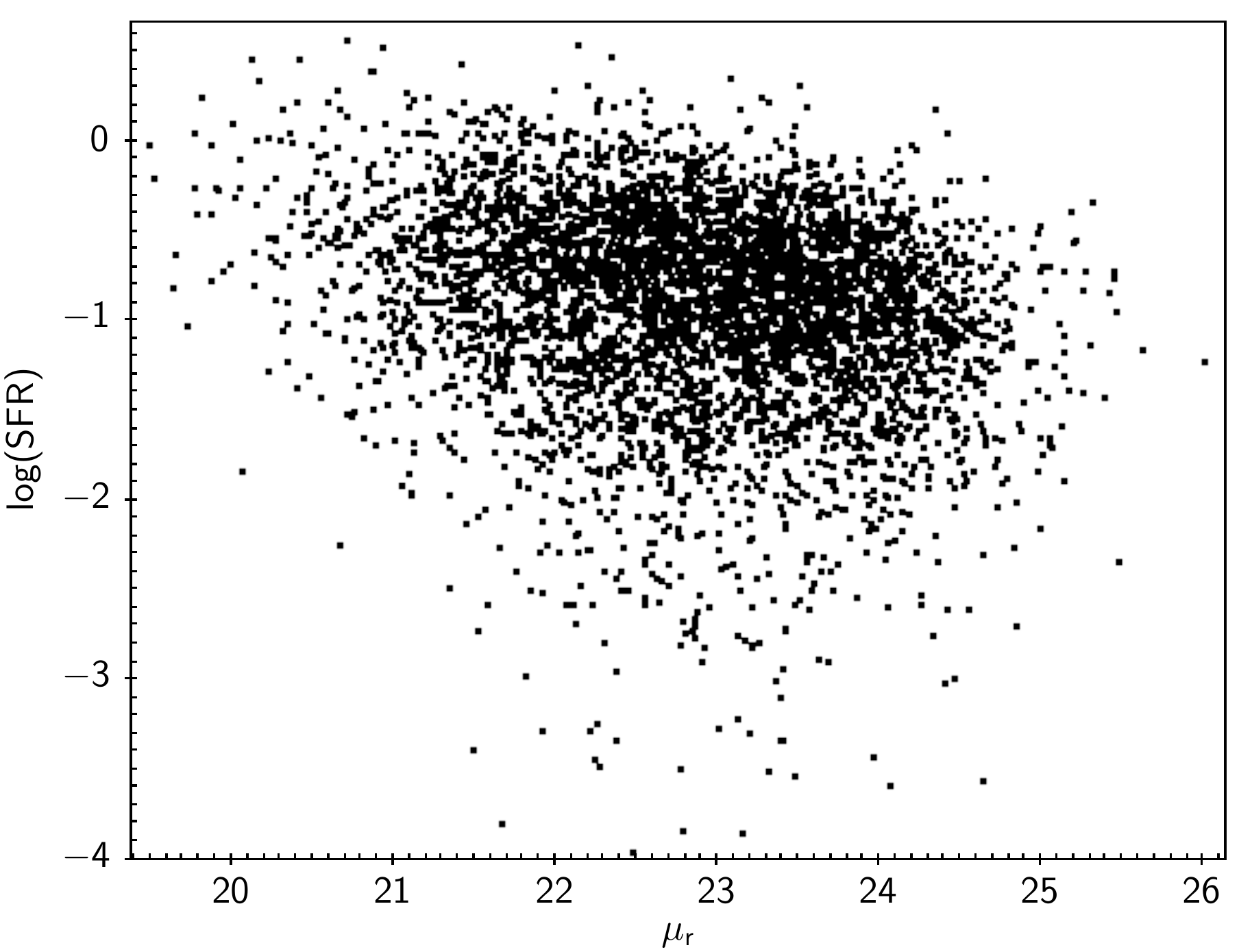}
\caption{Variation of SFR (in $M_{\odot}$/yr) with $\mu_r$ (in magnitudes per square arc second) for Sample 1.
}
\label{onesersicsbsfr}
\end{figure}

\begin{figure}
\includegraphics[width=\linewidth]{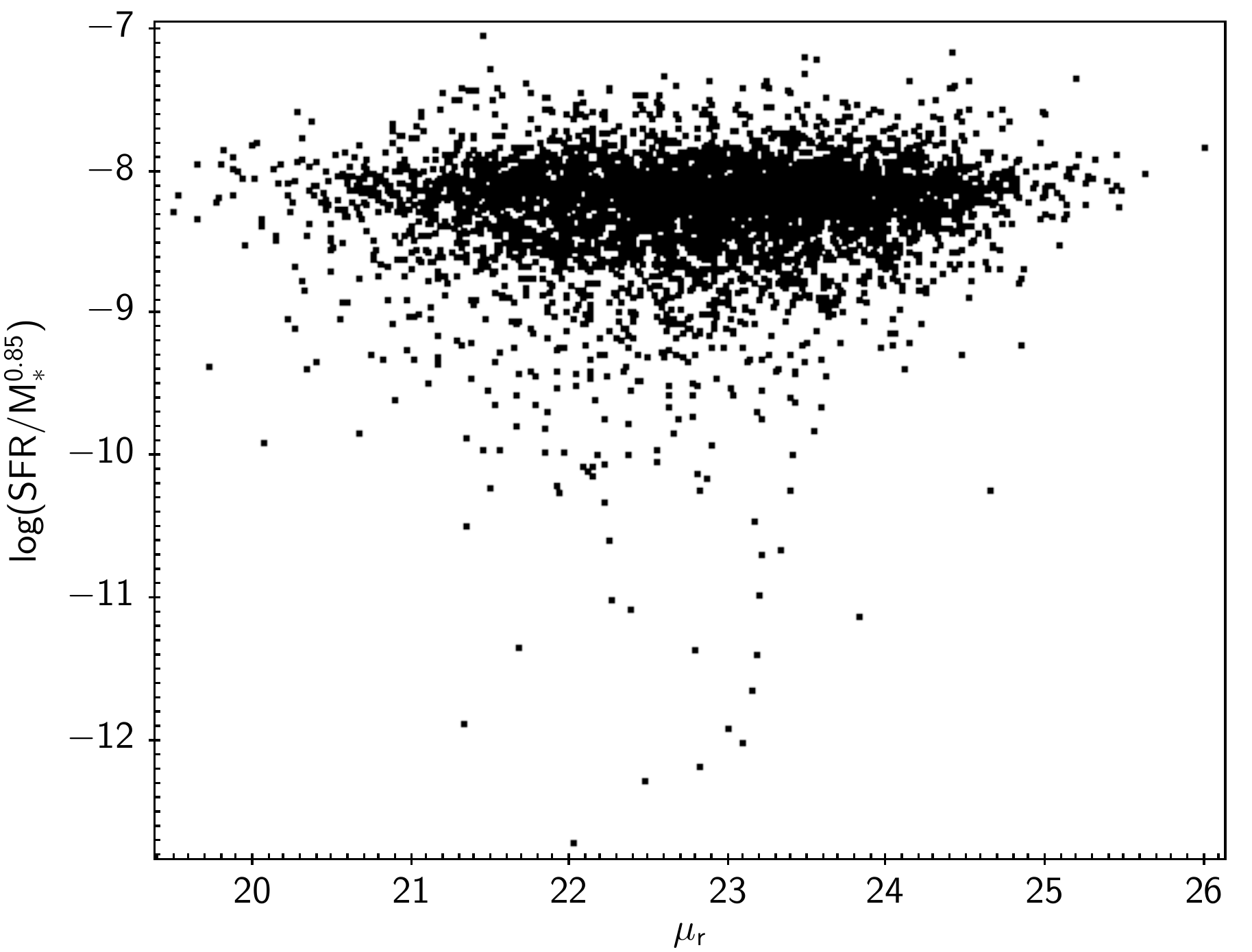}
\caption{As Figure 6 but for the mass-corrected star formation rate parameter m$_1$SFR (in units of $M_{\odot}^{0.15}$/yr).
}
\label{onesersicsbmsfr1}
\end{figure}

We can now search for any remaining dependence on SB.
Fig. \ref{onesersicsbsfr} shows the overall dependence of SFR on $\mu_r$, while Fig. \ref{onesersicsbmsfr1} shows the equivalent plot for the mass-corrected m$_1$SFR. It is evident that while the SFR does vary with SB \citep[e.g..][]{Phillipps1985, BelldJ2000, HunterElm, Oneil2007}, once the primary mass dependence of the SFR is removed, there is no significant remaining dependence of SFR on SB (Pearson correlation coefficient $r=0.09$), at least for galaxies in our sampled luminosity range. Note that \cite{Bell2012} found essentially the same result for low redshift galaxies if we assume that $U-V$ colour (their figure 7) is a proxy for the mass normalised SFR.

\begin{figure}
\includegraphics[width=\linewidth]{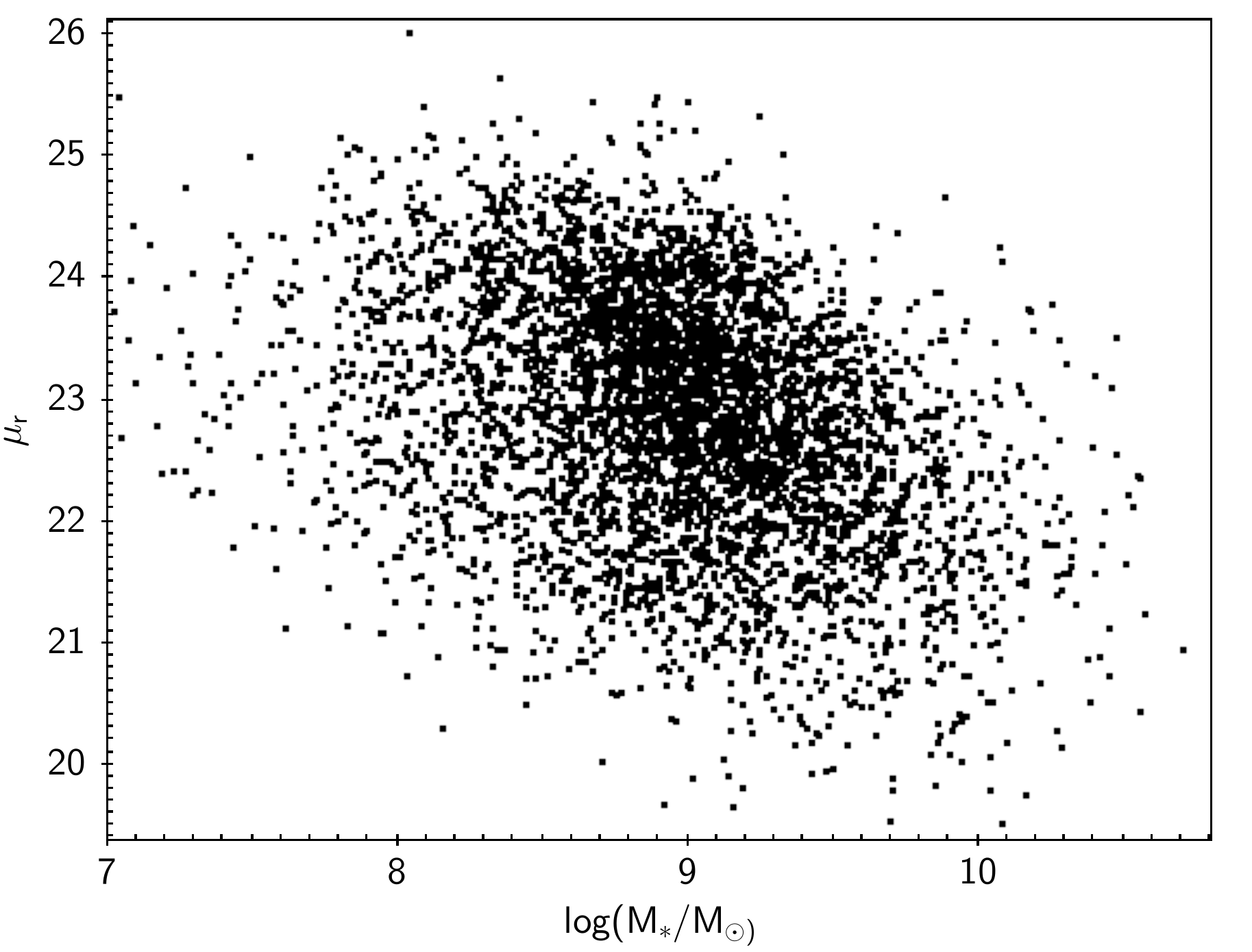}
\caption{Variation of surface brightness (in magnitudes per square arc second) with stellar mass (in solar masses) for Sample 1.
}
\label{onesersicmasssb}
\end{figure}

One might, of course, argue that since SB is correlated with $M_*$, the primary dependence of SFR could be the one on mass density, as in Fig. 6, rather than mass. However as is evident from Fig. 5, the correlation between SFR and mass is much tighter. The relation between SB and mass is actually also a rather broad one (see Fig. \ref{onesersicmasssb}).

\begin{figure}
\includegraphics[width=\linewidth]{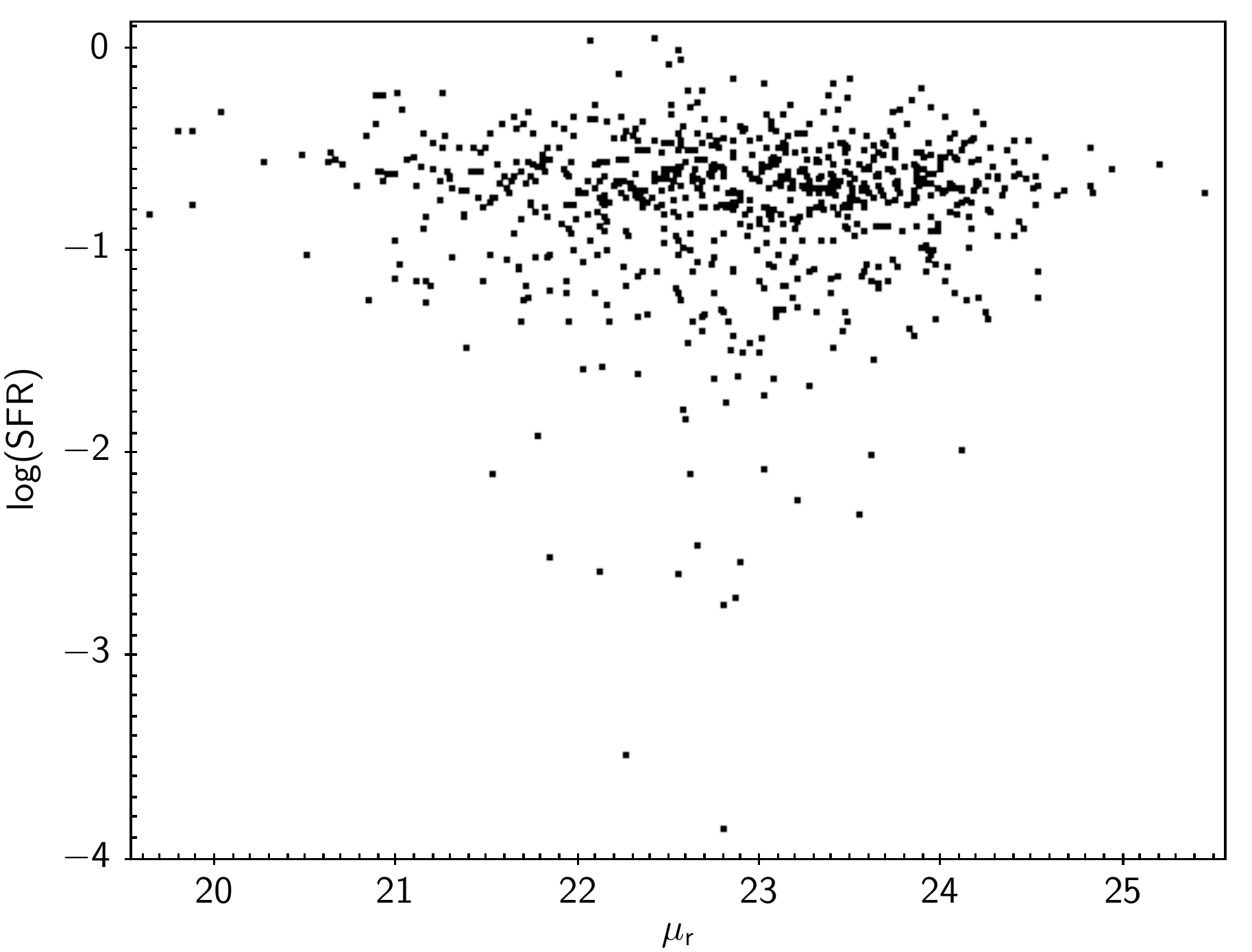}
\caption{Variation of SFR (in $M_{\odot}$/yr) with $\mu_r$ (in magnitudes per square arc second) for the single-component galaxies with mass between $10^9$ and $10^{9.2} M_{\odot}$.
}
\label{onesersicsbsfr90}
\end{figure}

We can further confirm the lack of a relationship between SFR and SB by plotting SFR versus SB at a specific mass, say around the sample's mass distribution peak at $10^9 M_{\odot}$, as shown in Fig. \ref{onesersicsbsfr90}; again no correlation is seen (Pearson $r = 0.01$). As a further check, we find that splitting the sample into `round' (fairly face-on) and `flat' (more edge-on), above and below an axis ratio $b/a = 0.5$, or using the $i-$band SB, makes no difference to the (lack of) correlation ($r < 0.1$ in all cases).

In these plots it is evident that there is a very wide range of SFR at a given SB, with neither the significantly star forming nor the quiescent galaxies showing a SB dependence. The errors in $\mu_r$ ($\sim 0.05$ mgnitudes per square arc second) are very much smaller than would be required to blur out a real correlation.

Next, we repeat the analysis for the discs from the two-component fits, Sample 2. (Recall that there are fewer objects in this sample). We first determine the mass dependence of the SFR in the same way as before. To obtain the disc mass we have simply scaled the overall mass by the disc fraction in $r-$band luminosity (see Fig. 2), without attempting to allow for any differences in M/L between the components. The mass-corrected SFR, m$_2$SFR, is defined similarly to the previous m$_1$SFR, though with a shallower mass dependence, $m_2=0.45$, as empirically found for the `main sequence' in this sample. (m$_2$SFR therefore has units of $M_{\odot}^{0.55}$/yr. Errors in m$_2$SFR are typically 0.08 dex).

\begin{figure}
\includegraphics[width=\linewidth]{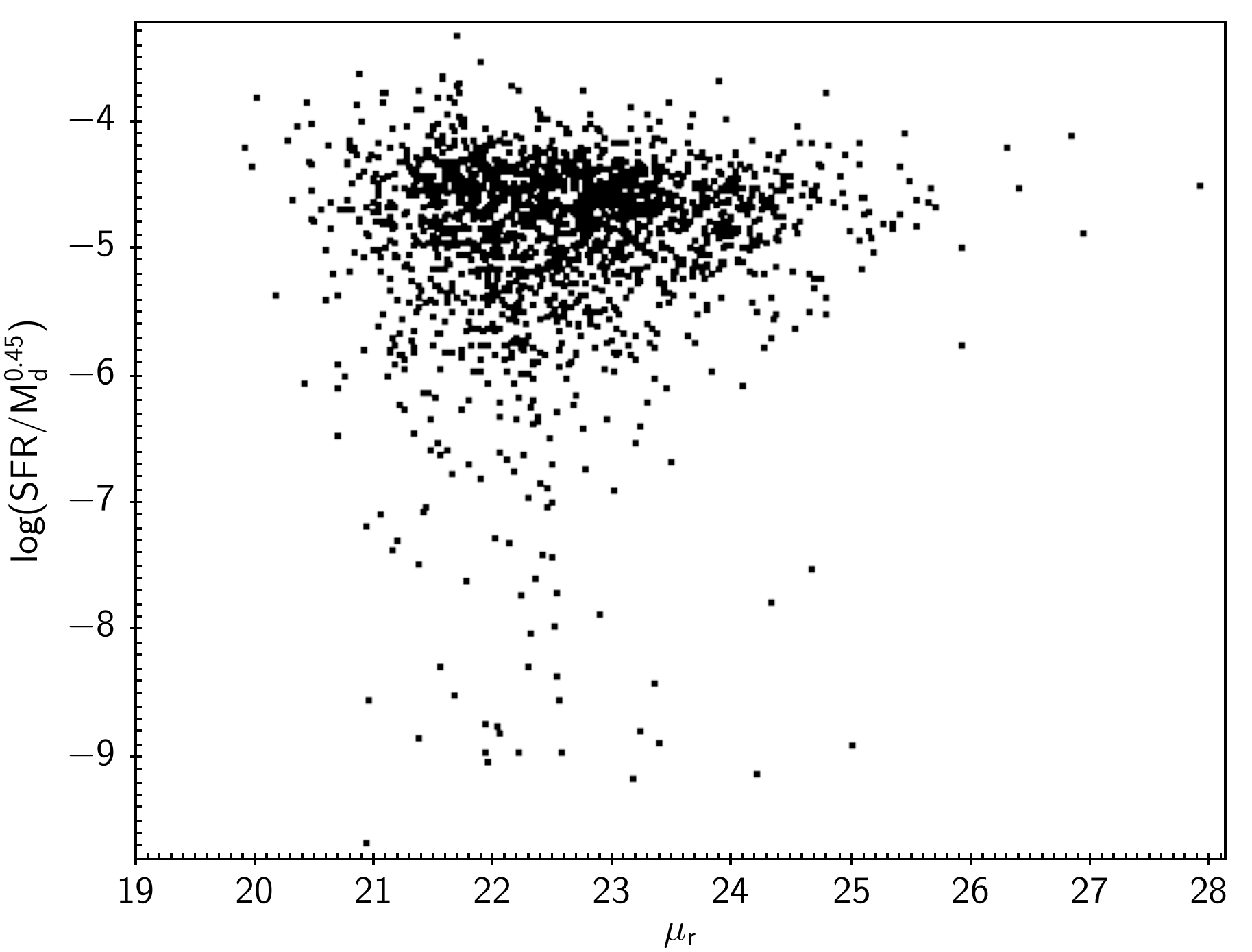}
\caption{Variation of mass-corrected star formation rate parameter m$_2$SFR (in units of $M_{\odot}^{0.55}$/yr) with disc surface brightness (in magnitudes per square arc second) for Sample 2.
}
\label{twosersicsbmsfr}
\end{figure}

\begin{figure}
\includegraphics[width=\linewidth]{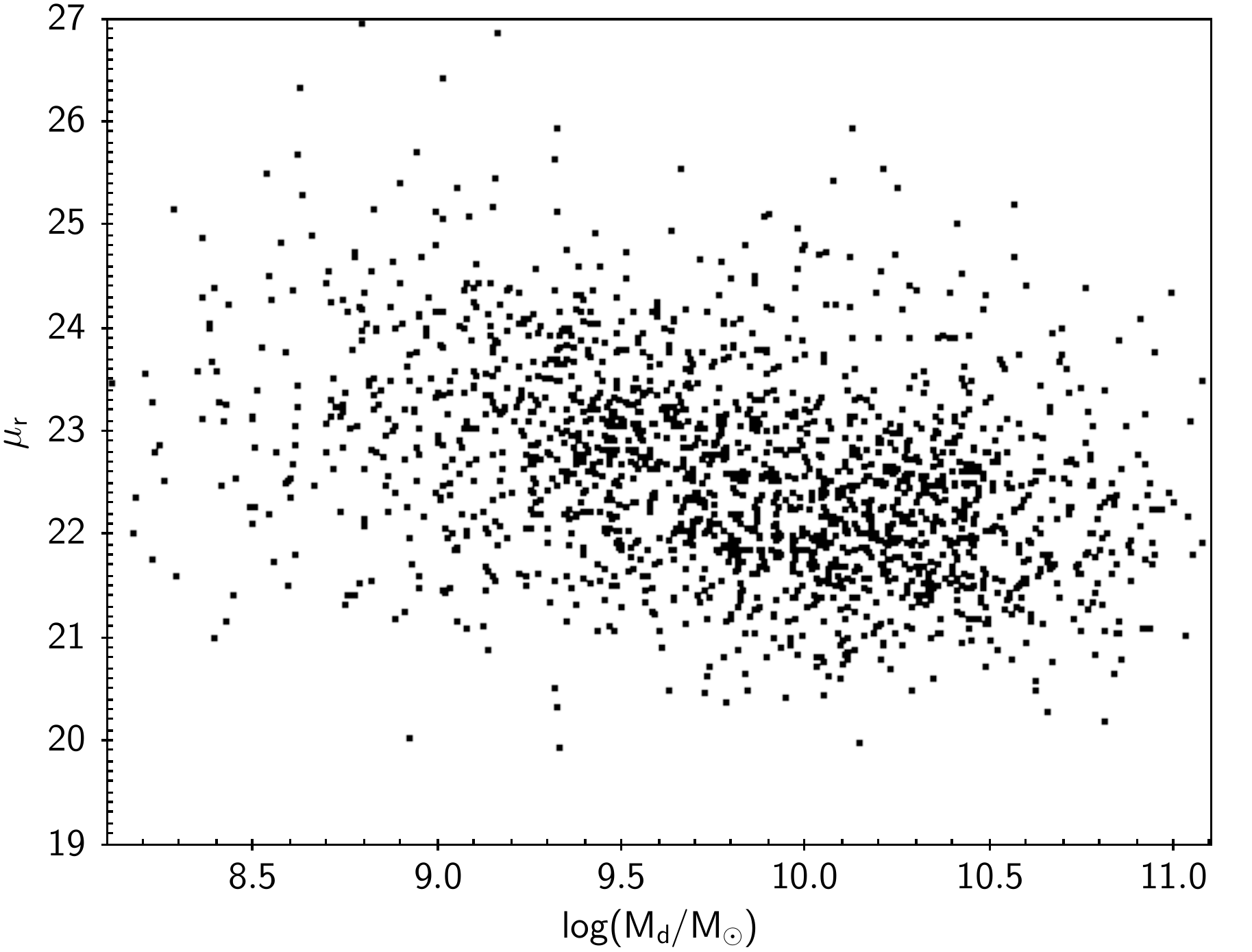}
\caption{Variation of disc surface brightness (in magnitudes per square arc second) with disc mass (in solar masses) for Sample 2.
}
\label{twosersicsbm}
\end{figure}

Fig. \ref{twosersicsbmsfr} shows the distribution of m$_2$SFR against $\mu_r$ for these disc components. It is evident that again there is no systematic trend of m$_2$SFR with SB ($r=0.05$), though the scatter increases towards the high SB end, creating a rather wedge shaped distribution. In particular, as noted above, the quiescent galaxies largely have high SB discs. As for Sample 1, they are also primarily of high mass. (For completeness, Fig. \ref{twosersicsbm} shows the overall distribution of SB with disc mass for Sample 2). Again, `round' and `flat' discs show the same lack of correlation. Similarly, the same distribution is seen for galaxies with resolved or with point-like `bulges'.

\subsection{Star Formation Rate Surface Density}

 It is, of course, well established that the surface density of SFR in galaxies, $\Sigma_{sfr}$, depends on their {\it gas} surface densities, $\Sigma_{gas}$ \citep{Schmidt1959, Kennicutt1998}, though with a substantial scatter.  In fact, a similar relationship, $\Sigma_{sfr} \propto \Sigma_{gas}^N$ with $N \simeq 1.4$, occurs within individual galaxies \citep[as seen in the radial profiles; e.g.][]{Kennicutt1989,Bigiel2008,Leroy2008}, as well as between them. \cite{Barrera2021} have recently summarised evidence that {\em within} individual galaxies $\Sigma_{sfr}$ correlates with both $\Sigma_{gas}$ (specifically the molecular mass) and the stellar surface density $\Sigma_*$.
 
\begin{figure}
\includegraphics[width=\linewidth]{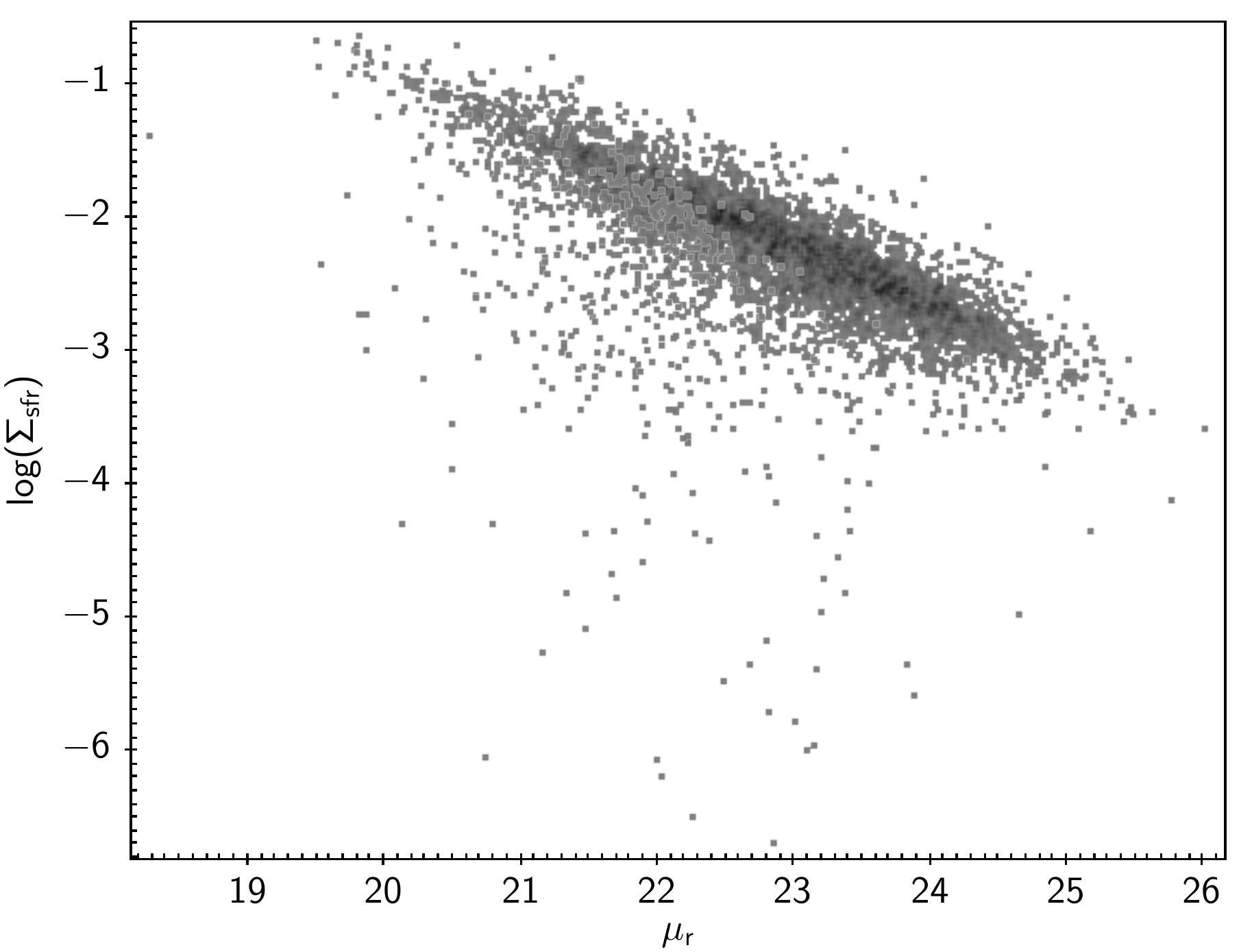}
\caption{Variation of surface density of star formation, $\Sigma_{sfr}$ (in units of $M_{\odot}$/yr/kpc$^2$) with surface brightness (in magnitudes per square arc second) for Sample 1.
}
\label{onesersicsbsfra}
\end{figure}

\begin{figure}
\includegraphics[width=\linewidth]{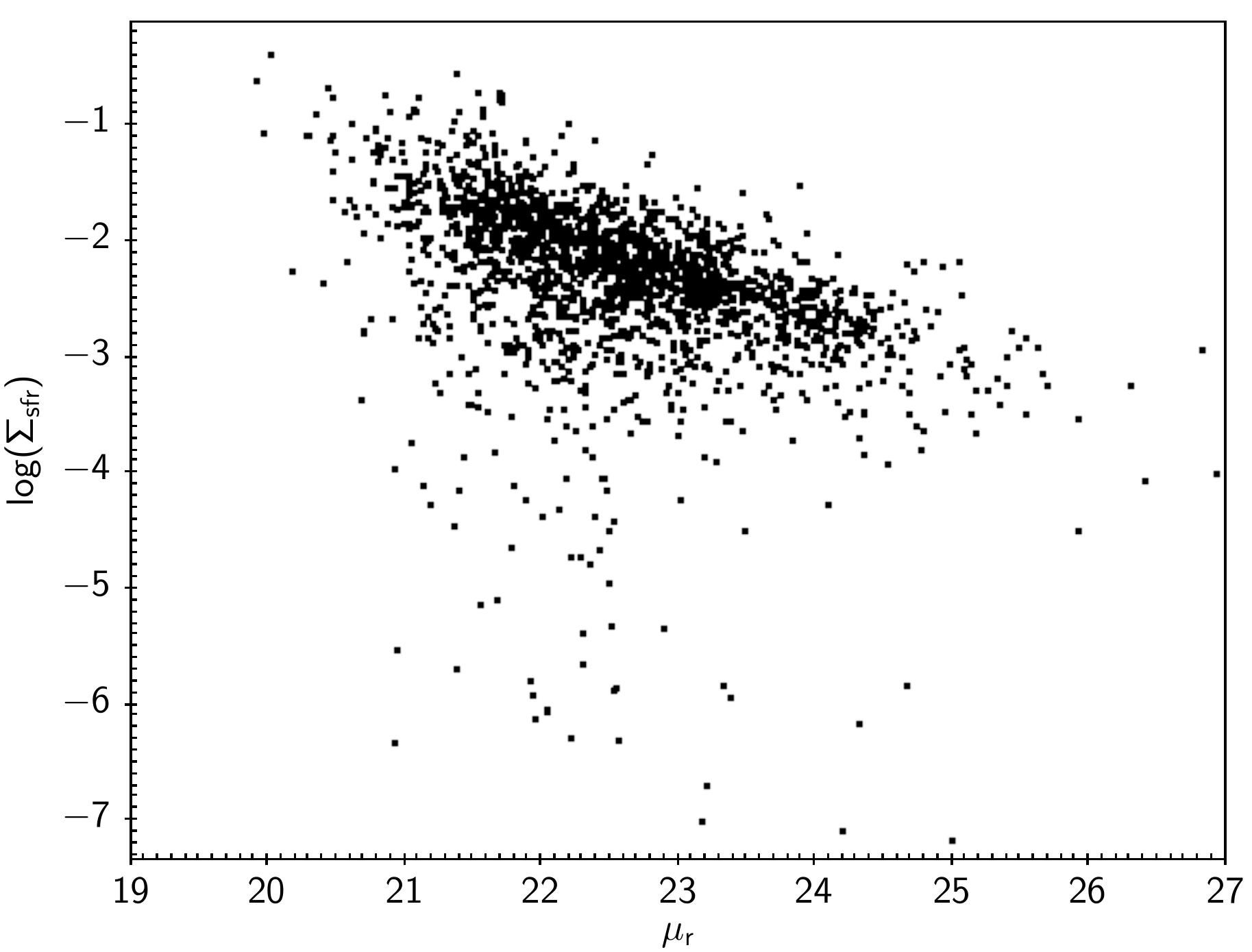}
\caption{Variation of surface density of star formation $\Sigma_{sfr}$ (in units of $M_{\odot}$/yr/kpc$^2$) with disc surface brightness (in magnitudes per square arc second) for Sample 2.
}
\label{twosersicsbsfra}
\end{figure}

 One might therefore consider global correlations between galaxies of their SFR surface density with their characteristic $r-$band SB (again as proxy for the corresponding stellar mass surface density). Fig. \ref{onesersicsbsfra} and Fig. \ref{twosersicsbsfra} show the variation of $\Sigma_{sfr}$ with $\mu_r$ for the one-component fits and the discs of the two-component fits, respectively. In each case the characteristic $\Sigma_{sfr}$, in units of $M_{\odot}$/yr/kpc$^2$, is calculated simply from the total SFR and the area corresponding to the effective radius. Errors in $\Sigma_{sfr}$ are typically 0.07 dex.
 
 This time we see very clear correlations between the SFR and SB. However, given the existence of a nearly linear correlation between SFR and luminosity (via the mass) for `main sequence' galaxies, dividing both SFR and $L$ by (the same) area to obtain $\Sigma_{sfr}$ and SB essentially guarantees a relation such as seen in Fig. \ref{onesersicsbsfra}. The less tight relation in Fig. \ref{twosersicsbsfra} follows because the  SFR-$L$ relation is not linear for Sample 2 discs, so dividing each by area does not simply move points parallel to the relation. Even so, the correlations seen are clearly largely forced by the SFR-$M_*$ relations and the use of the same areas when calculating $\Sigma_{sfr}$ and $\mu_r$.

\subsection{Gas and Dust}
 
\begin{figure}
\includegraphics[width=\linewidth]{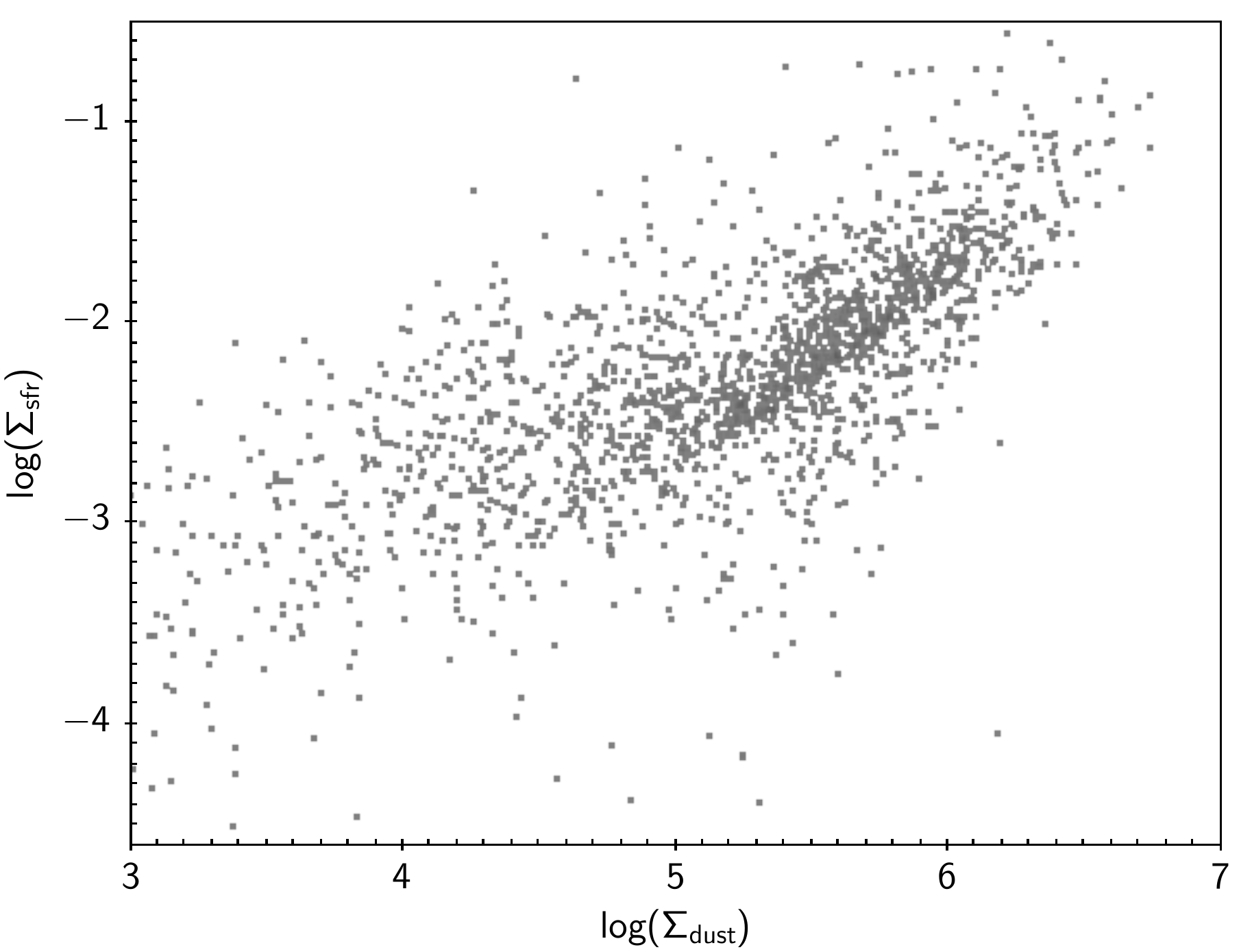}
\caption{Variation of surface density of star formation $\Sigma_{sfr}$ (in units of $M_{\odot}$/yr/kpc$^2$) as a function of dust mass surface density $\Sigma_{dust}$ (in $M_{\odot}$/kpc$^2$), for discs of Sample 2.
}
\label{twosersicdustsfr}
\end{figure}

 As an aside, although we do not have gas densities in the GAMA data, it is of interest to look at the dust content as a proxy \citep{Eales2010}, and indeed in its own right since correlations between SFR and dust mass are well established \citep{daCunha2008,daCunha2010}. In addition, \cite{Dalcanton2004} and \cite{Holwerda2019} suggest that the existence of dust lanes is related to disc gravitational instability. Fig. \ref{twosersicdustsfr} shows (for Sample 2) $\Sigma_{sfr}$, as above in units of $M_{\odot}$/yr/kpc$^{2}$, versus the dust surface density $\Sigma_{dust}$ in $M_{\odot}$/kpc$^2$, the latter calculated in the same way from the total dust mass $M_{dust}$ (as provided by ProSpect  with typical errors of 20\%, 0.08 dex) and the effective radius. We can see that we retrieve a strong, near-linear correlation (as did \cite{Grootes2013}, who also used $M_{dust}/R_e^2$ as a measure of $\Sigma_{dust}$ and $M_*/R_e^2$ for $\Sigma_*$, from a detailed analysis of a small sample of GAMA galaxies). However, this is again expected because of the near-linear SFR-$M_{dust}$ relation in \cite{daCunha2010} \citep[see also][]{Rowlands2014, Beeston2018}, for the same reason as for the $\Sigma_{sfr} - \mu_r$ relation in the previous section.

\begin{figure}
\includegraphics[width=\linewidth]{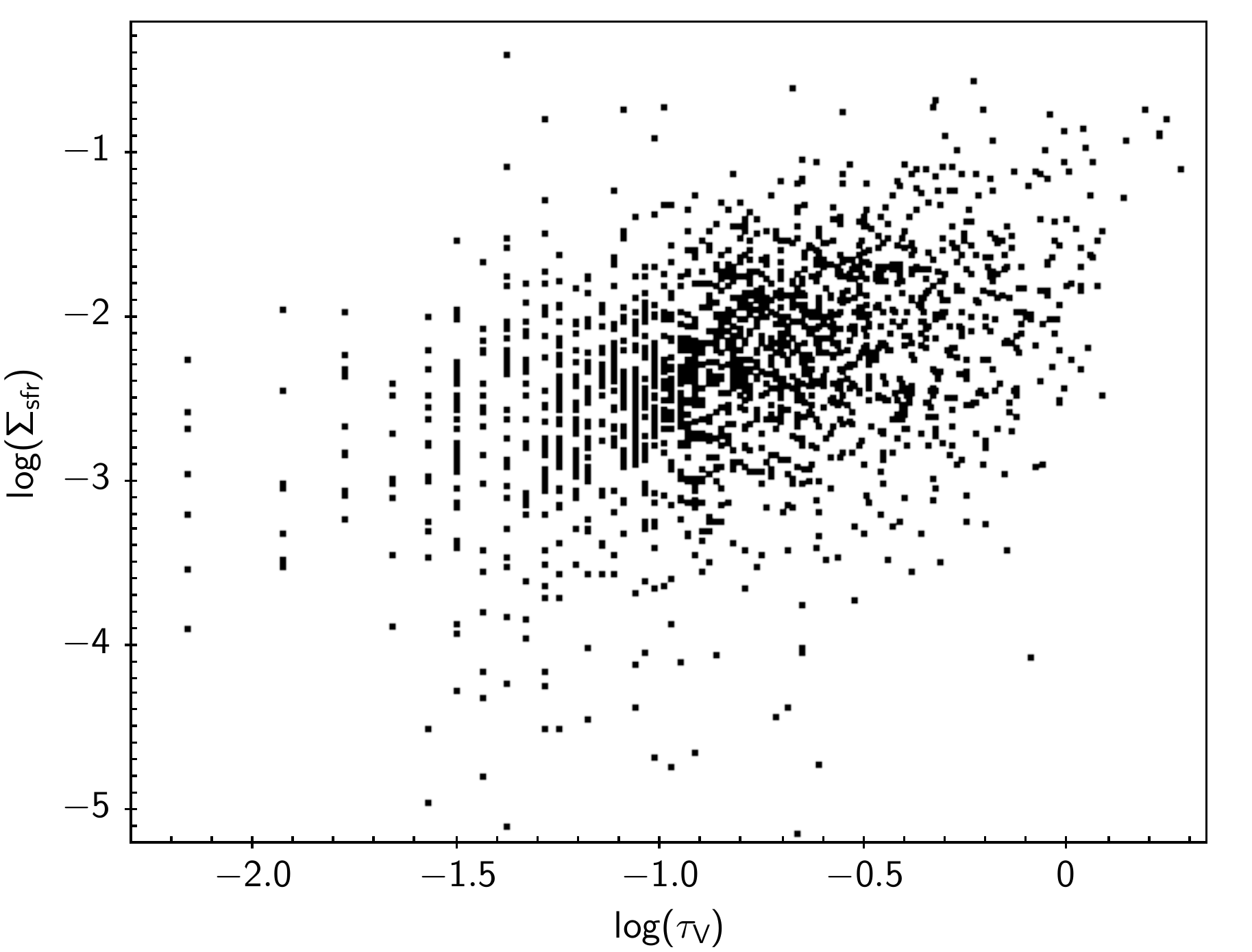}
\caption{Variation of surface density of star formation, $\Sigma_{sfr}$, in units of $M_{\odot}$/yr/kpc$^2$, as a function of dust optical depth in the $V$-band, $\tau_v$, for discs of Sample 2.
}
\label{twosersicdepthsfr}
\end{figure}

In this case, though, we can check whether there is a real correlation underlying Fig. \ref{twosersicdustsfr} since we can alternatively plot the $V-$ band disc optical depth $\tau_v$ derived from the MAGPHYS SED fitting (with typical errors of 20\%). This also represents the (diffuse) dust disc surface density but does not involve division by the area. Fig. \ref{twosersicdepthsfr} confirms that $\Sigma_{sfr}$ does indeed vary with the dust surface density in this representation (and by presumption the gas surface density), as we would expect. \footnote{The vertical `stripes' at low log($\tau_v$) are due to the discrete values of $\tau_v$ in the MAGPHYS inputs.} Though with a range of 2 dex at given $\tau_v$, the plot is broadly compatible with a relation, $\Sigma_{sfr} \propto \tau_v$ for the significantly star-forming (`main sequence') galaxies. 

We should not, of course, expect there to be a perfect correspondence between the measured dust and the star-forming gas, not least because of the generally differently distributed atomic and molecular components, but if we assume that the relationship between SFR and gas available for fuel is the fundamental one, then this correlation with $\Sigma_{dust}$ implies that random variations in gas-to-dust ratio are small enough to keep the correlation intact, even though this ratio is known to itself vary with mass and a galaxy's SFH \citep{DeVis2017b}, which will affect the slope we measure.

\begin{figure}
\includegraphics[width=\linewidth]{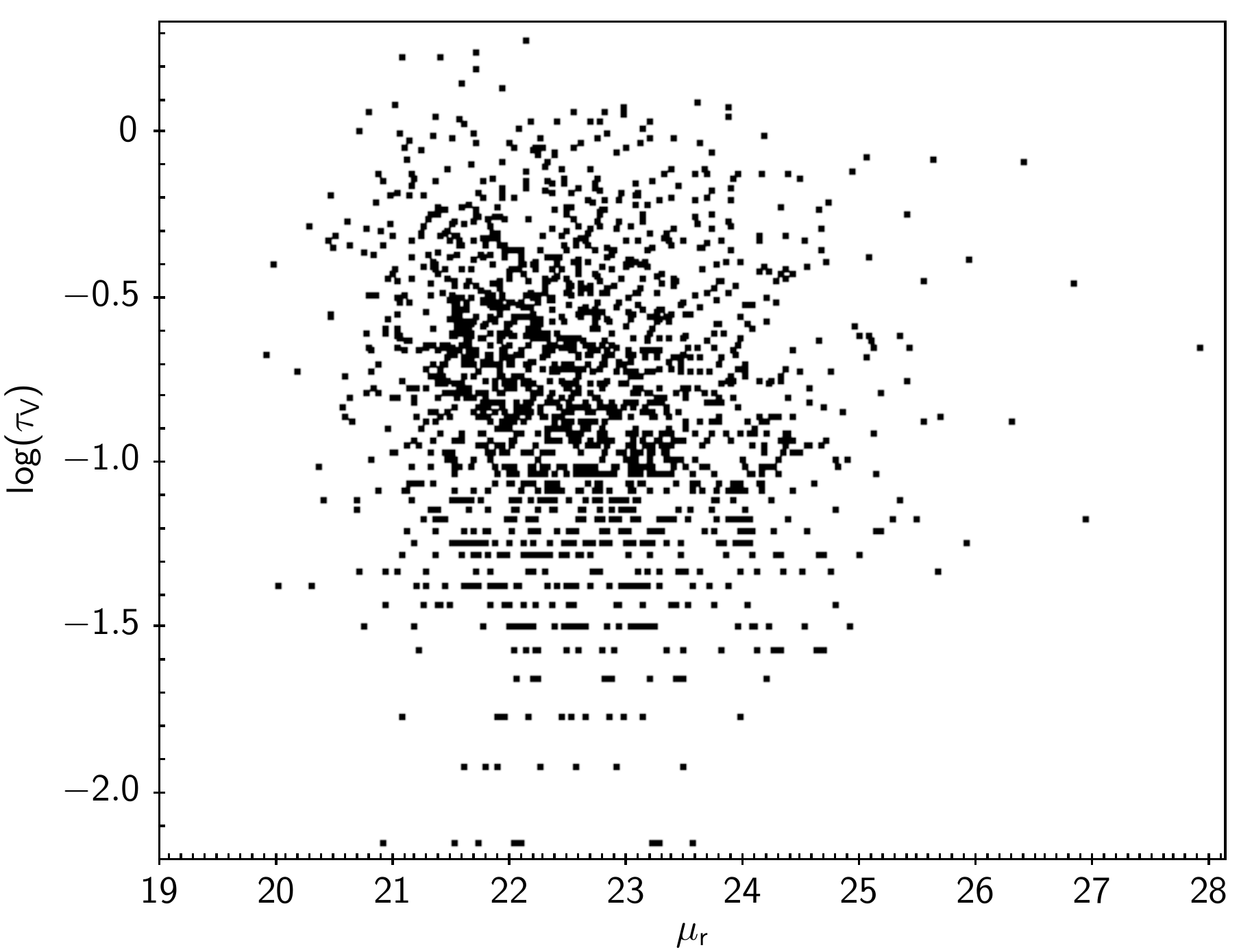}
\caption{Variation of dust optical depth in the $V-$band, $\tau_v$, as a function of surface brightness (in magnitudes per square arc second) for discs of Sample 2.
}
\label{twosersicdepthsb}
\end{figure}

Fig. \ref{twosersicdepthsb} then shows the dust optical depth $\tau_v$ against the surface brightness $\mu_r$. This shows that characteristic surface brightness does not track the variation of (mean) surface density of dust (and gas) between galaxies (Pearson $r=0.09$), consistent with its lack of correlation with SFR. \cite{DeVis2017a} show a similar lack of a significant relationship between $\Sigma_*$ and ultra-violet extinction, if their separate group of very gas-rich, low-luminosity LSBGs is excluded. 

\section{Discussion}

What, then, should we make of the perhaps surprising main result above, that for galaxies of any given total stellar mass, the star formation rate (and also the surface density of star formation) does not depend on the $r-$band surface brightness, and by inference the stellar mass surface density? There are a number of ways of looking at this. 

For instance, it is clear that if we take a set of (proto-)disc galaxies with different initial surface densities of gas and let each evolve to similar remaining gas fractions (say a typical value for present day spirals, around 10-20\%), then those with high stellar mass density will also have high (remaining) gas surface density and by implication high SFR, clearly counter to what is seen.

However, the non-linear slope of the Schmidt/Kennicutt relation between gas density and SFR implies that high gas density systems convert a larger fraction of their gas into stars per unit time. Thus we can imagine that the initially high density systems have more quickly reduced their gas content and therefore fuel supply, so that, after a certain evolutionary time, their SFR is no higher than that of their initially lower density counterparts.

There would seem to be two testable implications of this. The high initial density systems should, in this picture, clearly have high current stellar surface densities, and hence SB, and should also show (a) large average stellar population ages because of the high initial SFR \cite[cf.][]{MacArthur2004}, and (b) low gas fractions because of the rapid gas depletion \cite[cf.][]{Bell2003}. Although, as above, the latter is not directly observable within our GAMA data, we might expect a corresponding low dust mass fraction.

\begin{figure}
\includegraphics[width=\linewidth]{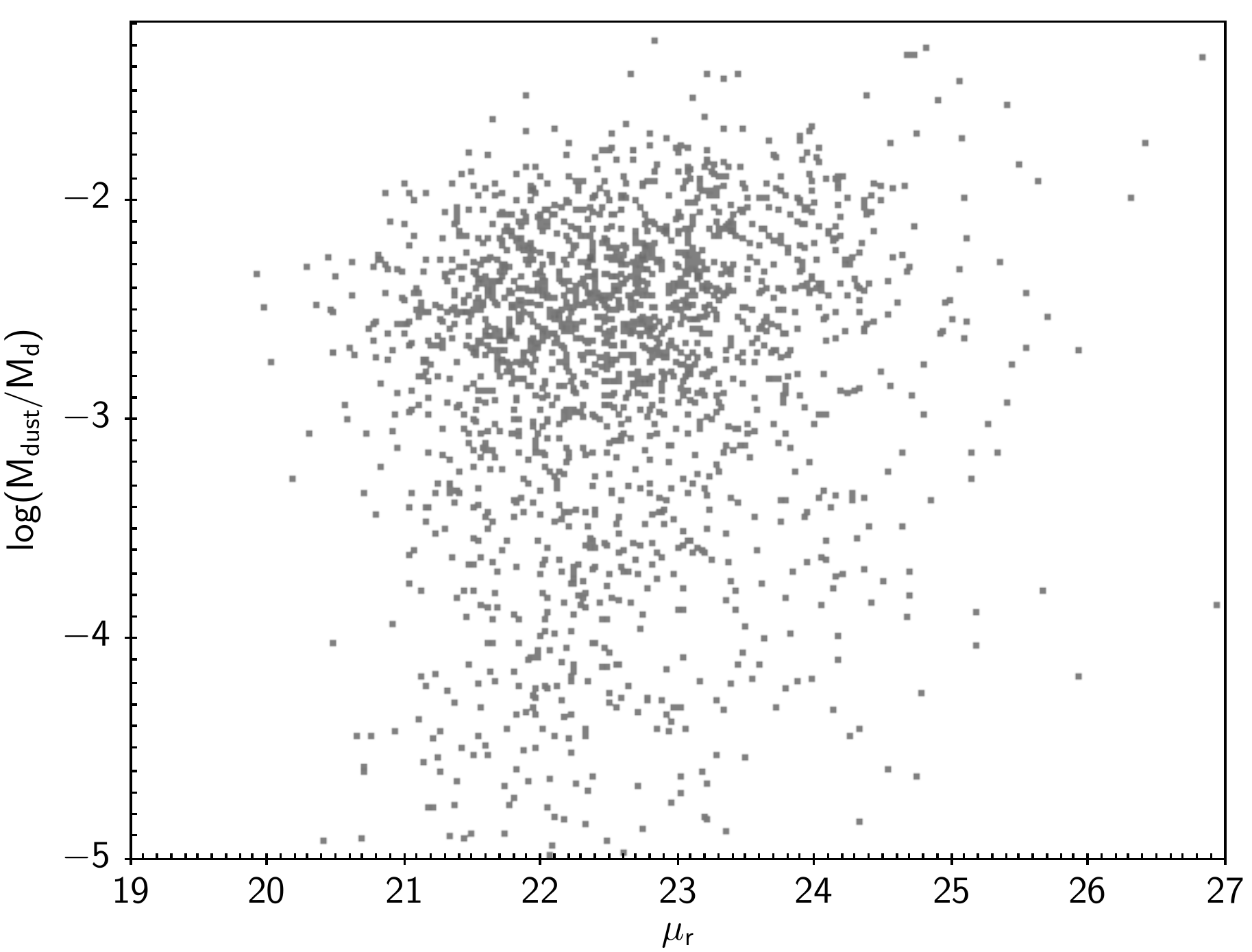}
\caption{Variation of the ratio of dust mass to stellar disc mass with disc surface brightness (in magnitudes per square arc second) for Sample 2.
}
\label{twosersicsbdust}
\end{figure}

Prediction (a) has already been shown to hold for GAMA galaxies; \cite{Robotham2022} find that at given total mass, large galaxies (i.e. those with low mass surface density) do have younger stellar ages than smaller (higher surface density) galaxies. For prediction (b), Fig. \ref{twosersicsbdust} shows the distribution of dust mass as a fraction of stellar disc mass against SB.  (Errors in $M_{dust}/M_d$ are around 0.13 dex). The galaxies with low dust mass fractions - and by implication low gas fractions - do tend to have high rather than low SB, but any overall trend is very weak  ($r=0.15$), high SB galaxies having a very wide range of dust-to-stellar mass ratios. Note that, although the outlier quiescent galaxies at low SFR are more likely to have low dust fraction, relative to the overall sample, removing them does not alter the appearance of Fig. 17 (i.e. it does not remove a significant number of the low dust fraction galaxies), nor does it significantly alter the correlation coefficient.

Thus we find only partial support for a toy model where the lack of dependence of the current SFR on stellar surface density is due to more rapid evolution of the high density systems cancelling out the expected variation with original gas surface density.

An alternative way to look at this is to consider specifically the late evolution of galaxies of a given original (gas) surface density. Once the gas content has reduced to 10\%, for instance, any remaining star formation can only change the stellar surface density by a small factor (corresponding to $\sim 0.1$ magnitudes in SB). However, as the gas is used up and the gas content reduces to, say, 1\%, then by the Kennicutt relations we would expect the SFR to reduce by a factor $\sim 30$. Thus if we observe such galaxies at a variety of evolutionary stages (hence gas fractions) we should expect a wide range of SFR at a particular SB, as seen in our data, and that this variation should correlate with gas fraction. 

\begin{figure}
\includegraphics[width=\linewidth]{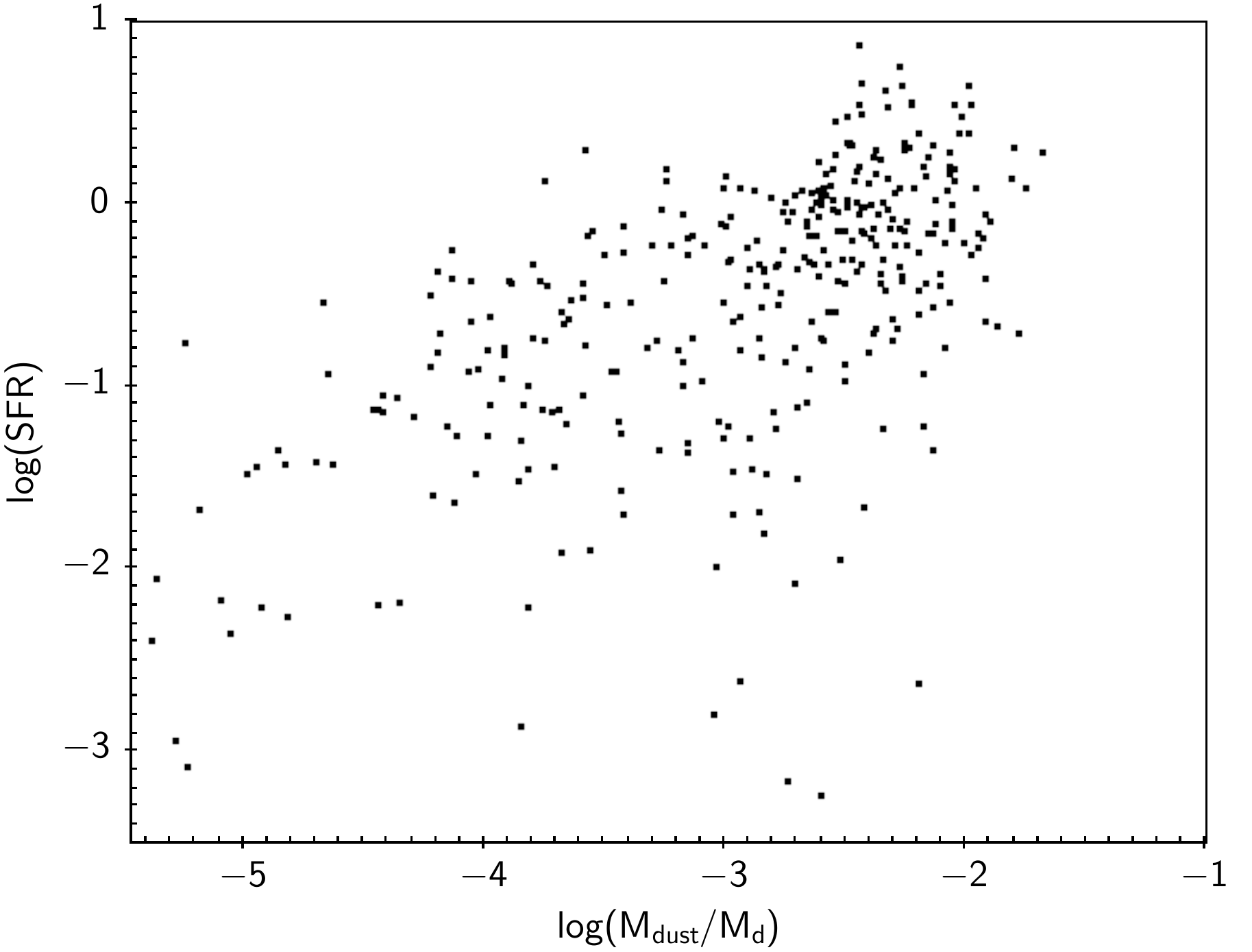}
\caption{Variation of SFR (in $M_{\odot}$/yr) with the ratio of dust mass to stellar disc mass for Sample 2 galaxies with disc SB limited to 22 - 22.5 magnitudes per square arc second.
}
\label{twosersic22dustsfr}
\end{figure}

To test this, galaxies have been selected from the two-component fit galaxies within the disc SB range $\mu_r = 22$ to 22.5 (the peak of the sample's SB distribution in Fig. 4). Fig. \ref{twosersic22dustsfr} shows the relation between the ratio of dust-to-stellar mass, as proxy for gas fraction as above, and SFR. We see the expected correlation, for this simple model, that at fixed SB (and thus fixed {\em original} gas density), SFR increases with gas fraction. If we select out other SB ranges, the points entirely overlap with those for  $\mu_r \simeq 22$,  

We must therefore conclude that at fixed mass, the pre-existing stellar surface density (and therefore the original total surface density of gas, barring any mergers) does not in anyway modulate the current SFR.

\section*{Acknowledgements}

GAMA is a joint European-Australasian project based around a
spectroscopic campaign using the Anglo-Australian Telescope.
The GAMA input catalogue is based on data taken from the Sloan Digital Sky Survey and the UKIRT Infrared Deep Sky Survey. Complementary imaging of the GAMA regions is being obtained by a number of independent survey programmes including GALEX MIS, VST KiDS, VISTA VIKING, WISE, Herschel-ATLAS, GMRT and ASKAP providing UV to radio coverage. GAMA was funded by the
STFC (UK), the ARC (Australia), the AAO, and the participating institutions. The GAMA website is http://www.
gama-survey.org/ . This paper has made extensive use of the TOPCAT software package \citep{MBTaylor}.

\section*{Data Availability}
The data underlying this paper are available at http://www.gama-survey.org/dr4/.

\end{document}